\documentclass[preprint,12pt]{elsarticle}
\usepackage{graphicx}
\usepackage[utf8]{inputenc}
\usepackage[english]{babel}
\usepackage{amsmath}
\usepackage{amsfonts}
\usepackage{amssymb}
\usepackage[left=2cm,right=2cm,top=2cm,bottom=2cm]{geometry}
\begin{document}
	\begin{frontmatter}
	\title{Transient Superdiffusive Motion on a Disordered Ratchet Potential\\}
	\author{D.G. Zarlenga}
	\address{Instituto de Investigaciones Cient\'ificas y Tecnol\'ogicas en Electr\'onica (ICYTE),\\
		Facultad de Ingenier\'{\i}a,\\
		Universidad Nacional de  Mar del Plata,\\
		Av. J.B. Justo 4302, \\
		7600 Mar del Plata, Argentina}
	\author{G.L. Frontini}
	\address{Instituto de Investigaciones en Ciencia y Tecnología de Materiales (INTEMA),\\
		Facultad de Ingenier\'{\i}a,\\
		Universidad Nacional de  Mar del Plata,\\
		Av. J.B. Justo 4302, \\
		7600 Mar del Plata, Argentina}
	\author{Fereydoon Family}
	\address{Department of Physics, Emory University, Atlanta,
		GA 30322,  USA}
	\author{C.M. Arizmendi}
	\address{Instituto de Investigaciones Cient\'ificas y Tecnol\'ogicas en Electr\'onica (ICYTE),\\
		Facultad de Ingenier\'{\i}a,\\
		Universidad Nacional de  Mar del Plata,\\
		Av. J.B. Justo 4302, \\
		7600 Mar del Plata, Argentina}
	\date{\today }

	\begin{abstract}
	The relationship between anomalous superdiffusive behavior and particle trapping probability is analyzed on a rocking ratchet potential with spatially correlated weak disorder. The trapping probability density is shown, analytically and numerically, to have an exponential form as a function of space. The trapping processes with a low or no thermal noise are only transient, but they
	can last much longer than the characteristic time scale of the system and therefore might be detected
	experimentally. Using the result for the trapping probability we obtain an analytical expression for the number of wells where a given number of particles are trapped.  We have also obtained an analytical approximation for the second-moment of the particle distribution function $C_2$ as a function of time, when trapped particles coexist with constant velocity untrapped particles. We also use the expression for $C_2$ to characterize the anomalous superdiffusive motion in the absence of thermal noise for the transient time.
\end{abstract}
\end{frontmatter}
	\section{INTRODUCTION}
	
	Even a small amount of disorder on periodic potentials can cause a rich phenomenology of transport and diffusive behavior.
Among the more impressive effects  is the increase in the diffusion coefficient  by several orders of magnitude
\cite{Khoury2011,Reimann2008}. This behavior
has been observed experimentally in the motion
of colloidal spheres through a periodic potential  \cite{Lee2006}.
Another important effect is anomalous diffusion in the form of  subdiffusion and superdiffusion in correlated disordered tilted periodic potentials \cite{Khoury2011}.
The anomalies in diffusion have been studied mainly in weakly disordered tilted periodic potentials
\cite{Khoury2011,Reimann2008}. Nevertheless, disorder may also appear in ratchets. The asymmetric potential characteristic of ratchets often arises in systems where structural bottlenecks lead to
 quenched disorder \cite{Chou2004}. Other examples of disorder on ratchets include heterosubstrate-induced graphene
superlattices in which zero-energy
states emerge in the form of Dirac points in asymmetric potentials \cite{Fan2016},
and antisymmetric dc voltage drop  along amorphous indium oxide wires exposed to an ac bias source \cite{Poran2011}.
When the thermal noise
is negligible, the overdamped equation of motion
accounts solely for effects of quenched disorder  \cite{denisov2007,Denisov2009} and
may be used for the study of the dynamics of localized structures
like  domain walls, driven vortices in type-II superconductors and driven Wigner crystals \cite{Olson1998, Pardo1998, Hellerqvist1996, Troyanovski1999, Reichhardt2001}.

Diffusion enhancement and superdiffusion has been obtained recently for a weakly disordered ratchet potential, in the absence of thermal noise \cite{Zarlenga2007}.
A trapping mechanism is shown to appear  whenever the quenched disorder strength is
higher than a threshold value with coexistence of locked and running states producing both  anomalous superdiffusive behavior and several orders of magnitude growth of the diffusion coefficient.

This pronounced enhancement over free thermal diffusion has been predicted theoretically \cite{Reimann2008} and observed experimentally \cite{Lee2006} in weak disorder tilted potentials.   Correlated spatial disorder exists in many disordered materials, such as polymers \cite{Cioroianu2016}, porous materials \cite{Adler1992}, and glasses
\cite{Berthier2011}.
Long-range correlated spatial disorder has also been found to produce anomalous diffusion in an overdamped rocking ratchet without thermal noise   \cite{Gao2003}.

Since the anomalous effects of  disorder appear at the transition from locked to running
solutions \cite{Khoury2011,Reimann2008}, in this work we focus on particle motion on a rocking ratchet potential with spatially correlated weak disorder  \cite{Popescu2000, Gao2003, Zarlenga2007}. Specifically, we concentrate on the particle trapping probability density, namely the probability density of existence of locked states, and its opposite behavior, the existence of running states.
Although
 trapping processes without  thermal noise have only transient nature, their dynamical behavior
could last much longer than the characteristic time scale of the system and therefore be detectable
experimentally. In fact, transient anomalous behavior has been predicted theoretically and observed experimentally \cite{Bronstein2009, Hanes2012, Hanes2013}.
In a recent very interesting work, a transient initial superdiffusive behavior followed by subdiffusion was found in ac-driven inertial Brownian ratchets, by numerical simulation and experimentally with an asymmetric
SQUID subjected to an external ac current and a constant magnetic flux \cite{Spiechowicz2015, Spiechowicz2016}.

The outline of the paper is as follows. In Sec. II we present the model and its associated dynamical equations. In Sec. III we analytically derive an exponential form for the trapping probability density and  compare it to the results of the numerical simulation of the dynamical equations.  In Sec. IV we consider the same problem, but ask a different question, namely, what is the effect of disorder on the trapping probability?  We also derive an analytical expression using the exponential form of the trapping probability. In Sec. V  we characterize the anomalous diffusive motion using the second moment as a function of time when the velocity of non-trapped particles is  constant. Finally, in Sec. VI we conclude with some comments and conclusions.

\section{MODEL}
We consider the overdamped motion of a particle in a ratchet potential $U(x)$ in the presence of a quenched noise, $\eta(x)$, with a Gaussian probability density function. The equation of motion is given by,
\begin{equation} \label{ratchet equation}
 - \gamma \dfrac{dx}{dt} - \big(1-\sigma\big) \dfrac{dU}{dx} - \sigma \dfrac{d\eta}{dx} + A\sin(\omega t)=0,
\end{equation}
where $\gamma$ is the friction coefficient, $\sigma$ is a measure of the relative strength of the ratchet potential compared to the quenched disorder. $A$ and  $\omega$ are the amplitude and the frequency of the sinusoidal driving force, respectively.

The deterministic ratchet-potential is represented by the sum of two sine functions with height $V$ and wavelength $\lambda$ as,
\begin{equation} \label{deterministic potential equation}
U(x) = - V\dfrac{\lambda}{2\pi} \bigg[ \sin\Big(\dfrac{2\pi x}{\lambda}\Big) + {\frac{1}{4}} \sin\Big(\dfrac{4\pi x}{\lambda}\Big) \bigg].
\end{equation}
The  spatial autocorrelation function of the ratchet potential has a maximum at:
\begin{equation} \label{deterministic autocorrelation at x equals 0}
G_{U}(0)= < U(x) U(x+0) > = \dfrac{1}{\lambda} \int_{0}^{\lambda} U^2(x)\,dx =  1.0625 \dfrac{V^2}{2} \dfrac{\lambda^2}{(2 \pi)^2}.
\end{equation}
In order to make the strengths of the ratchet force and the quenched disorder comparable, we express the quenched disorder spatial autocorrelation in the form,
\begin{equation} \label{quenched disorder autocorrelation}
G_{\eta}(x)= 1.0625 \dfrac{V^2}{2}\dfrac{\lambda^2}{(2 \pi)^2} \enskip e^{-\dfrac{(2 \pi)^2 x^2}{2 l^2}},
\end{equation}
where the length $ l$ is defined such that $\Lambda= l/\lambda$ is the width of the Gaussian distribution in dimensionless spatial dimension $z$, which we will define in the next paragraph.

Instead of using arbitrary values of model parameters, we normalize them by defining the dimensionless spatial variable $z$ and the dimensionless temporal variable $\tau$ as:  $z=2 \pi x / \lambda$ and  $\tau=[(2 \pi)^2 V/ (\gamma \lambda^2)] t$.  Using these parameters, the dimensionless form of the dynamical equation becomes,
\begin{equation} \label{normalized ratchet equation}
- \dfrac{dz}{d \tau} - \big(1-\sigma\big) \dfrac{dU'(z)}{dz} - \sigma \dfrac{d\eta'(z)}{dz} + \Gamma sin(\Omega \tau) =0,
\end{equation}
where $\Gamma={A \lambda}/{(2 \pi V)}$, $\Omega={\omega \gamma \lambda^2}/{[(2 \pi)^2 V]}$, $U'(z)=U(z)/V$, and $\eta'(z)=\eta(z)/V$.

The quenched-disorder force autocorrelation can be determined from the random potential autocorrelation:
\begin{equation} \label{autocorrelation of derivative}
G_{F}=-\dfrac{d^2G_{\eta}}{dz^2}=-\dfrac{d^2}{dz^2} \bigg(\dfrac{1.0625}{2} e^{-\dfrac{z^2}{2 \Lambda^2}} \bigg)
\end{equation}
Then,
\begin{equation} \label{random force autocorrelation equation}
G_{F}=1.0625 \dfrac{1}{2 \Lambda^2}  \bigg( 1 - \dfrac{z^2}{\Lambda^2} \bigg)  e^{-\dfrac{z^2}{2 \Lambda^2}}
\end{equation}

To obtain the quenched disorder $\eta(z)$, we first generate values of random force at positions $x=0, x=\lambda_{RF}, x=2\lambda_{RF}...$. We interpolate the values for other positions, as needed, by using sync interpolation. To generate the random force values we use the method described in \cite{Rodriguez2007} with a finite-input-response filter (FIR filter) instead of an infinite-input-response filter (IIR filter) used in \cite{Rodriguez2007}. We use the FIR filter because FIR filters are always stable and  can be generated  faster than IIR filters.  The $\lambda_{RF}$ value is initially set to $2 \pi /100$, where $2 \pi$ is the ratchet spatial period. This is a starting value. Then it is checked to see if the number of filter taps to account for the main part of the autocorrelation function is at least $64$, so that a good autocorrelation function sampling is made. The main part of the autocorrelation function represents the range in $z$ for which the autocorrelation function value exceeds $1\%$ of the maximum autocorrelation function at $z=0$. In order to have at least $64$ filter taps, the range in $z$ must equal at least $(64 \enskip 2 \pi /100)$, initially. If the number of filter taps is less than $64$, the autocorrelation function is not properly sampled. Then, $\lambda_{RF}$ is reduced by a factor of ten. The reduction by a factor of ten is continued until the number of filter taps is at least $64$. As a test, we found that the disorder autocorrelation obtained with this method obeys Eq. \ref{random force autocorrelation equation}.

As seen from Eq. \ref{autocorrelation of derivative}, for a given $z$, a decrease in $\Lambda$ results in a decrease in the random-force potential autocorrelation. Thus, the potential changes abruptly , and its derivative, which is the random force, will develop high peaks that will trap the particles as they encounter regions of strong negative-force during their motion.
We define a 'well' as the distance between two consecutive peaks of the ratchet potential.  In dimensionless units, the range of $z$ values for the $n$th well is $(n-1)2\pi<z<n2\pi$. As in previous work \cite{Popescu2000}, we use the value $\Omega$=0.1, and define $v_{\omega}$ as the  speed of the particle that travels a well in a time period \~T$=2\pi / \Omega$.

\begin{figure}[t]
	\includegraphics[scale=0.3]{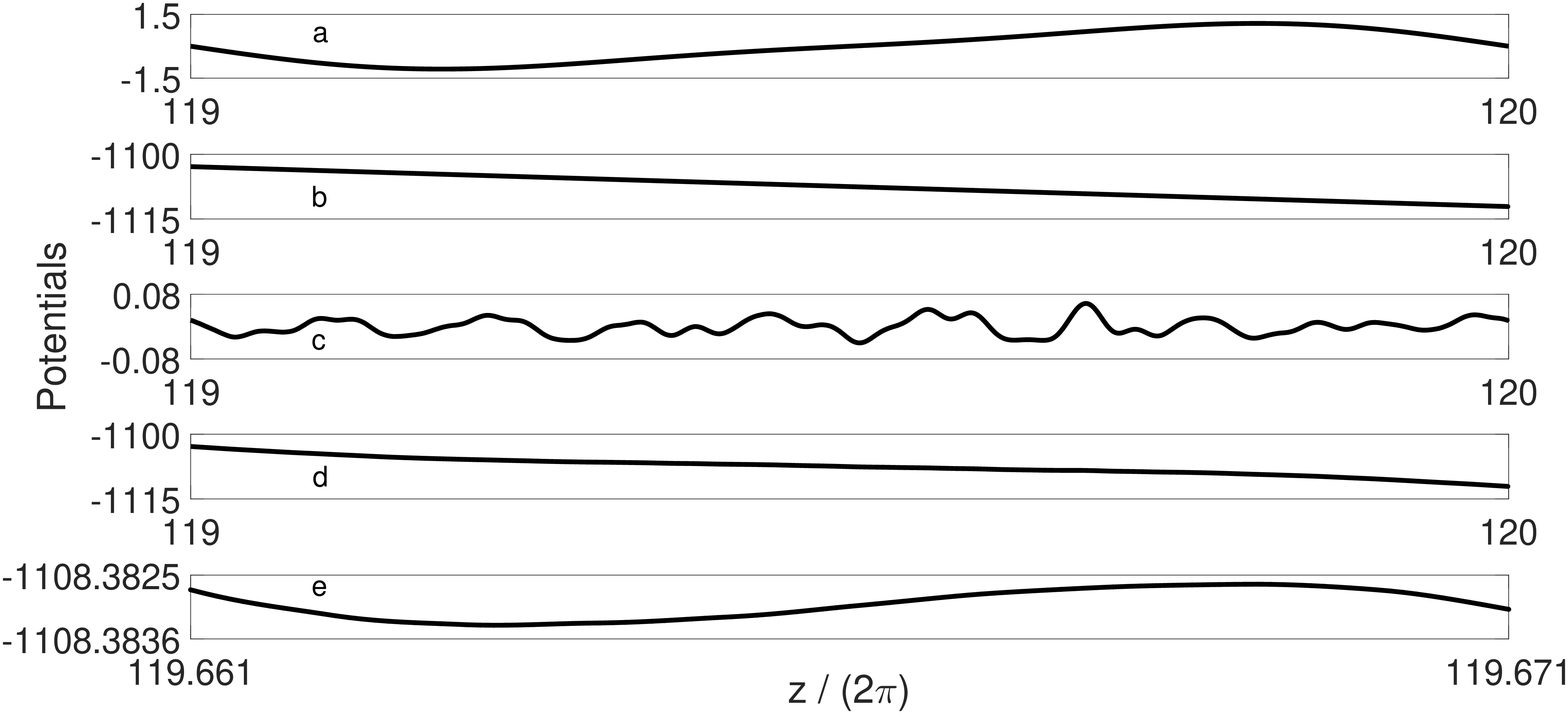}
	\caption{Potentials for $\Gamma=1.475$, $\Lambda=0.095$, and $\sigma=0.025$. (a) Asymmetric ratchet potential. (b) External-force potential which is a negative-derivative straight line. (c) Single-outcome quenched-disorder potential. (d) Resulting potential. (e) Zoom of (d) in which a very small region with positive derivative, called trap, is shown.}
	\label{ThePotentials}
\end{figure}

In order to show an example of particle trap, Fig. \ref{ThePotentials} shows all the potentials involved. The quenched-disorder potential is a single outcome for $\Gamma=1.475$, $\Lambda=0.095$, and $\sigma=0.025$, which are the same parameters as those in Fig. \ref{histogram2}. The external force is considered at the time of maximum positive value, since the particle will never go through the trap if it is not able to do it at this unbeatable condition. Figures (a) through (d) depict the potentials for the whole spatial period in which the trap appeared. Fig. (d) is the addition of figures (a) through (c). Fig. (e), however, is a zoom of Fig. (d), that allows us to note the very small positive-derivative region of the resulting potential. In this region, the resulting force is negative. The particle is unable to go through this region. Then, the region is called a trap.

\section{TRAPPING PROBABILITY DENSITY}

Let us suppose that the trapping of particles is characterized by a constant probability of trapping per unit length $\alpha_P$. We define  $\phi(z)$ as the probability that the particle is not trapped up to position $z$.  Then, $\phi(z)$ satisfies the relation,
\begin{equation} \label{conditional probability for trapping position1}
\phi(z+ \Delta z)=\phi  \big ( z+ \Delta z | \phi(z)=1 \big)  \phi(z)  +  \phi \big(z+ \Delta z  | \phi(z)=0 \big)  [1-\phi(z)],
\end{equation}
where $\phi(x|y)$ is the conditional probability of $x$ given $y$. Using the definition of $\alpha_P$, Eq. \ref{conditional probability for trapping position1} reduces to,
\begin{equation} \label{conditional probability for trapping position2}
\phi(z+ \Delta z)=(1-\alpha_P \Delta z)  \phi(z).
\end{equation}
This implies that the trapping probability $P(z)$ up to position $z$ satisfies the relation,
\begin{equation} \label{conditional probability for trapping position3}
\dfrac{dP(z)}{dz}+\alpha_P P(z) = \alpha_P
\end{equation}
\begin{equation} \label{trapping position4}
P(z) = 1 - e^{-\alpha_P z}
\end{equation}

 As a consequence,  the spatial trapping probability density may be approximated by the exponential function of the form,
\begin{equation} \label{pdf for trapping position}
f_{tr} (z)=\alpha_P  e^{- \alpha_P z },
\end{equation}

In order to check this result we have carried out simulations with different realizations of the quenched disorder. A number of  simulations  are carried out for each realization of the disordered potential, so that the statistical averages are over the number of  simulations.
Both the deterministic ratchet force and the external sinusoidal force are bounded, while the quenched disorder is not, because it has a Gaussian probability distribution.  As a result, in an infinitely-long path, every particle will, sooner or later, get trapped.

\begin{figure}
	\includegraphics[scale=0.3]{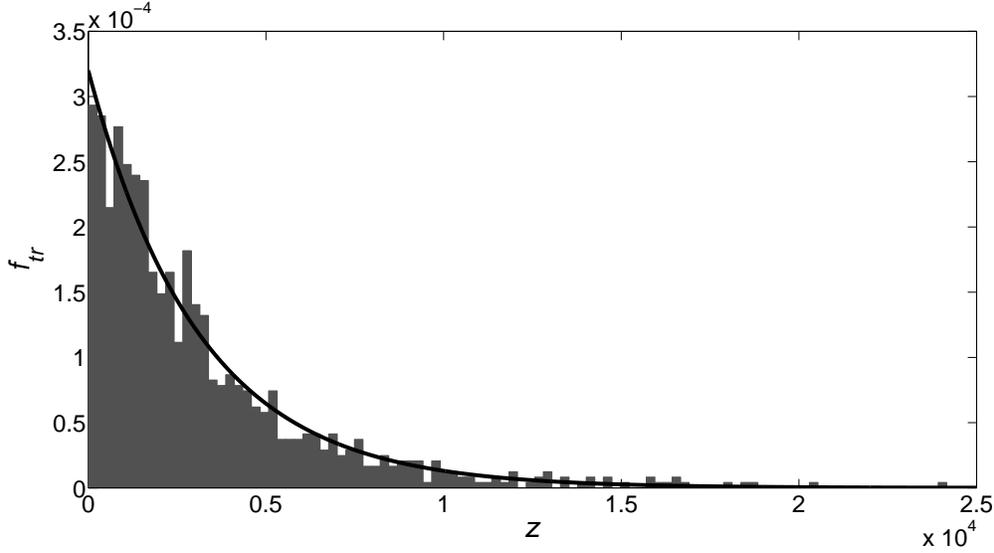}
	\caption{The spatial trapping probability $f_{tr}(z)$ is plotted as a function of distance $z$ for 1,000 simulations where the particle is  released at $z=0$. The solid line is an exponential fit to the data using Eq. \ref{pdf for  trapping position}.}
	\label{histogram2}
\end{figure}

The simulation results for the spatial trapping probability $f_{tr}(z)$
as a function of distance $z$ with $\Gamma=1.475$, $\Lambda=0.095$, and $\sigma=0.025$  is shown in Fig. \ref{histogram2}. The solid line is an exponential fit to the data.  There is excellent agreement with equation \ref{pdf for trapping position}.  We have found similar quality fits to the data for other values of  $\Gamma$ and $\Lambda$.

 The assumption that the trapping probability per unit length $\alpha_P$ does not depend on $z$ is based on the fact that  the disorder probability density function is the same for all $z$.  According to Eq. \ref{normalized ratchet equation}, a particle gets trapped when the quenched disorder (QD) at position $z$ is such that:
\begin{equation} \label{quenched disorder for trapping}
F_{QD} < - \bigg[ \frac{\Gamma}{\sigma} + \frac{1 - \sigma}{\sigma} F_R(z) \bigg],
\end{equation}
where $F_R(z)$ is the ratchet force. Then, if the disorder probability density function remains constant for all $z$, the probability that Eq. \ref{quenched disorder for trapping} is satisfied will be the same for all ratchet potential wells. Therefore, the trapping probability per well length will be the same for all wells. If long distances are considered, it can be assumed that $\alpha_P$ is a constant independent of $z$. In order to test that the disorder probability density function remains constant for all $z$, we carried out 5,000 simulations. It was found that the disorder probability density function was a Gaussian, with the same variance for all $z$. The same result was obtained for an exponentially-spatially-correlated disorder. In this way, the assumption that $\alpha_P$ is a constant is verified.\\
To take thermal noise into account, we add to Eq. \ref{ratchet equation} a noise term $\xi(t)$  that satisfies the autocorrelation function:
\begin{equation} \label{non-normalized thermal noise}
\big \langle \xi(t) \xi(t') \big \rangle = 2 \gamma k_B T \enskip \delta(t-t'),
\end{equation}
with strength $\gamma$. After normalization, the autocorrelation function reduces to:
\begin{equation} \label{normalized thermal noise}
\big \langle \epsilon(\tau) \epsilon(\tau ') \big \rangle = 2 \Theta \enskip \delta(\tau-\tau '),
\end{equation}
where
\begin{equation} \label{Theta value}
\Theta = \frac{k_B T}{V},
\end{equation}
and $\epsilon(\tau)$ is the dimensionless noise term that is added to Eq. \ref{normalized ratchet equation}.\\
When thermal noise is present, trapped particles are freed, and no particle is trapped forever. Then, Eq. \ref{pdf for trapping position} does not hold if we consider a particle at infinite time. However, for short times and low thermal noise Eq. \ref{pdf for trapping position} may be considered valid. In order to test this, trapped particles were analyzed at $\tau =2500$\~T for different $\Theta$ values. 
For low noise intensity, up to $\Theta=10^{-4}$ Eq. \ref{pdf for trapping position} still holds. For higher $\Theta$ values the trapping probability density is no longer an exponential. It can be said, as a consequence, that Eq. \ref{pdf for trapping position} is also valid for low thermal noise.
\begin{figure}[t]
	\includegraphics[scale=0.3]{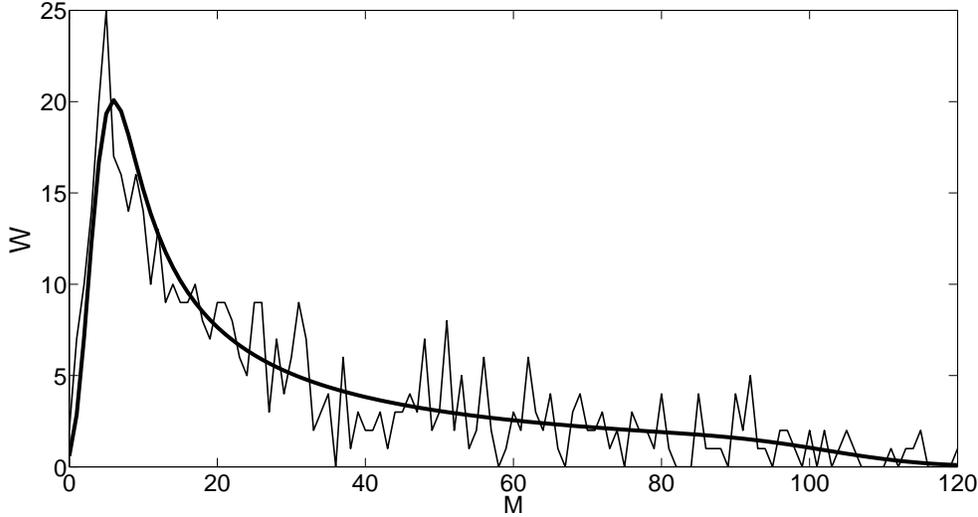}
	\caption{The prediction of Eq. \ref{NumberofWells} for the average number of wells $\overline{W(M)}$ where $M$ particles get trapped is shown as solid line. For comparison we have also plotted the simulation results for $N=16,000$ particles, with $Q=500$ wells from $z=0$ to $z=500$ with $\Gamma=1.5$, $\Lambda=0.085$, and $\sigma=0.025$ for which $\alpha_P=0.00104$.}
	\label{AmountOfWellsWithMParticles1b}
\end{figure}

\section{NUMBER OF WELLS WHERE A GIVEN NUMBER OF PARTICLES ARE TRAPPED}
We may wonder: if we simulate the behavior of $N$ particles, what is the most-likely number of wells in which $M$ out of $N$ particles will get trapped?  In order to answer this question we use Eq. \ref{pdf for trapping position} to calculate the cumulative distribution function,
\begin{equation} \label{cumulative distribution function}
F_{tr} (z)=1-e^{- \alpha_P z}.
\end{equation}
Using Eq. \ref{cumulative distribution function} we can calculate the probability that a particle gets trapped at well $n$ as:
\begin{equation} \label{probability of getting trapped at well n1}
P_n=e^{- \alpha_P 2\pi(n-1)}-e^{- \alpha_P 2\pi n}.
\end{equation}
As a result, the probability that a particle that started moving at $z$=0 arrives at well $n$ is given by,
\begin{equation} \label{1 minus cumulative distribution function}
1-F_{tr}\big( 2\pi(n-1)\big)=e^{- \alpha_P  2\pi(n-1)}.
\end{equation}
Therefore, the probability $p_n$ of getting trapped at {\it well} $n$ for an arriving particle is the ratio of the probability of getting trapped at well $n$ divided by the probability of arriving at well $n$:
\begin{equation} \label{probability of getting trapped at well n2}
p_n=\frac{P_n}{1-F_{tr}\big( 2\pi(n-1)\big)}=1-e^{-  2\pi\alpha_P}.
\end{equation}
Finally, the probability for an arriving particle to get trapped at {\it any well} is,
\begin{equation} \label{conditional probability of getting trapped at any well}
p_n=p=1-e^{-  2\pi \alpha_P }.
\end{equation}
If a large amount $N$ of particles evolve from $z=0$, the number of particles arriving at well $n$ will be  $K(n)= [N \exp(- 2\pi(n-1)\alpha_P )]$, and the chance that $M$ particles will get trapped at well $n$ is equal to,
\begin{equation} \label{probability that M particles get trapped at well n}
P_{M}(n)=\left( \begin{array}{c}
K(n) \\ M\end{array} \right) p^M (1-p)^{K(n)-M} .
\end{equation}
When $K(n)>>M$, we can approximate the Binomial distribution with the Poisson distribution:
\begin{equation} \label{Poisson approximation}
P_{M}(n)\approx \frac{e^{-K(n) p} \left\lbrace K(n) p\right\rbrace^M}{M!}
\end{equation}
If we consider the first $Q$ wells, the average number of wells where $M$ particles are trapped  is obtained as the sum of the probabilities $P_{M}(n)$ of getting trapped in each well, since independence of these events is assumed, namely:
\begin{eqnarray} \label{NumberofWells}
 \overline{W(M)} \approx \sum_{n=1}^{Q} \frac{e^{- K(n) p} \left\lbrace K(n) p\right\rbrace^M}{M!}
\end{eqnarray}

\begin{figure}[t]
	\includegraphics[scale=0.3]{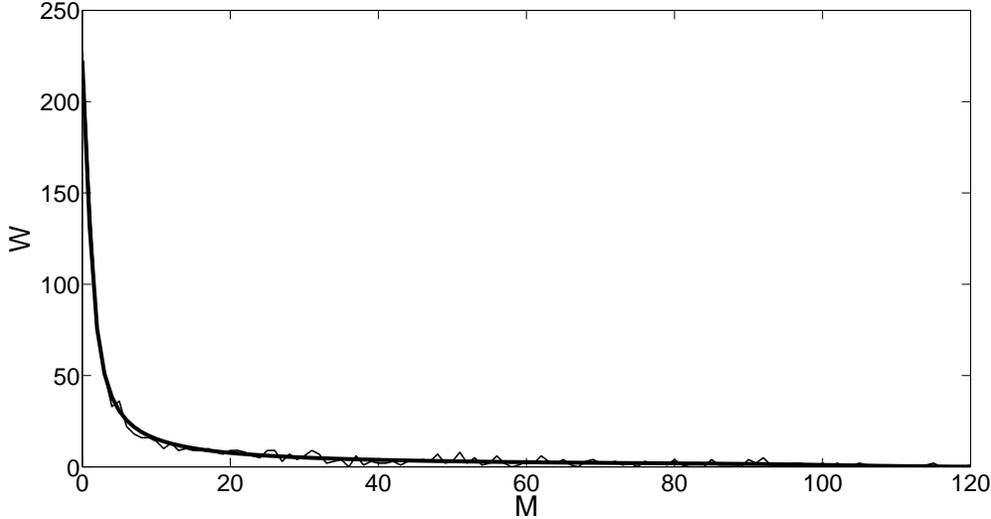}
	\caption{The same conditions as Fig. \ref{AmountOfWellsWithMParticles1b} but now with $Q=1,000$. As new far-away-from-the-origin wells are considered, more cases of wells with few trapped particles appear. The solid line is the prediction of Eq. \ref{NumberofWells} and gives an excellent fit to the  simulation data.}
	\label{AmountOfWellsWithMParticles2b}
\end{figure}

Eq. \ref{NumberofWells} is plotted as a solid line in Figs. \ref{AmountOfWellsWithMParticles1b} and \ref{AmountOfWellsWithMParticles2b} long with the simulation results for $N$=16,000 particles.  In Figs. \ref{AmountOfWellsWithMParticles1b} and \ref{AmountOfWellsWithMParticles2b}  values of $Q$ are 500 and 1,000, respectively.
\section{CHARACTERIZATION OF ANOMALOUS DIFFUSION}
To characterize the nature of the anomalous diffusion we now concentrate on the second moment $C_2(\tau)= \overline{z(\tau)^2} - {\overline{z(\tau)}}^2$. Let us assume that all particles were released at $\tau$=0, and $z=0$. After time  $\tau$ the particle has moved across $s$ wells and the particle density is given by,
\begin{equation} \label{particle density}
f(z)=\alpha_P e^{- \alpha_P z } \big[ u(z)-u(z-sv_\omega\tau) \big] +  \delta \big(z-sv_\omega\tau \big) \int_{sv_\omega\tau}^{\infty} \alpha_P e^{- \alpha_P z} dz,
\end{equation}
where $u(z)$ is the Heaviside step function. Eq. \ref{particle density} can be explained by assuming that particles that have not been trapped by time $\tau$ are all located at $z=sv_\omega\tau$. This is because the particles that have not been trapped have a single velocity value.  The average position can be calculated from Eq. \ref{particle density}:
\begin{equation} \label{average position one}
\overline{z(\tau)}= \int_{0}^{sv_\omega\tau} z \alpha_P  e^{- \alpha_P z} dz  +  (sv_\omega\tau)  \int_{sv_\omega\tau}^{\infty} \alpha_P  e^{- \alpha_P z }  dz=\dfrac{1- e^{-\alpha_T \tau}}{\alpha_P},
\end{equation}
where we have defined $\alpha_T=\alpha_P s v_\omega \tau$.  The mean of $z$ squared is then given by,
\begin{equation} \label{mean squared zee}
\overline{{z(\tau)}^2}= \int_{0}^{sv_\omega\tau} z^2  \alpha_P  e^{- \alpha_P z } dz  +  (sv_\omega\tau)^2  \int_{sv_\omega\tau}^{\infty} \alpha_P  e^{- \alpha_P z}  dz.
\end{equation}
Combining these results, we can calculate the second cumulant,
\begin{equation} \label{see sub too}
C_2 = \overline{z(\tau)^2} - {\overline{z(\tau)}}^2 =\dfrac{s^2  v_{\omega}^{2} } {\alpha_{T}^{2}} [1 - 2 \alpha_T \tau  e^{- \alpha_T \tau} - e^{ -2 \alpha_T \tau}] .
\end{equation}

\begin{figure}[t]
	\includegraphics[scale=0.3]{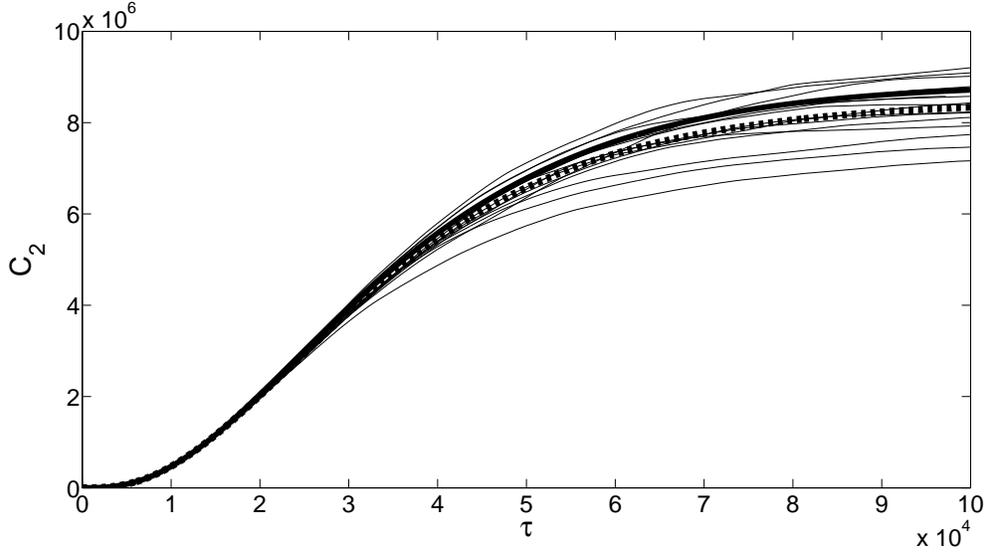}
	\caption{Second moment as a function of the normalized time. In this case $\Gamma=1.47$, $\lambda=0.095$. Each thin line was built by solving the differential equation for 1,000 particles. The thick dotted line is the average of the simulation data. The thick solid line is the prediction of  Eq. \ref{see sub too}, with $\alpha_P=0.000337$, $\alpha_T=0.0000672$ and $s=2$.}
	\label{C2_lambda0p095_Gamma1p47}
\end{figure}

\begin{figure}
	\includegraphics[scale=0.3]{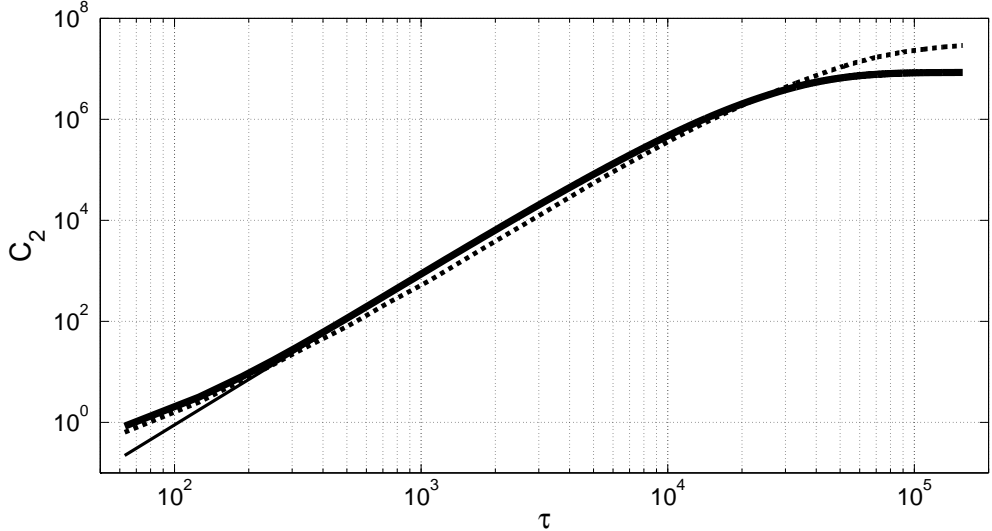}
	\caption{Log-log plot of $C_2$ versus $\tau$. The thick solid line was built out of simulations with 1,000 particles.  The thin solid line is the prediction of Eq. \ref{see sub too}. In early times the particle motion is ballistic, but over an extended range of time, both our analytic result and the simulations show that the motion is superdiffusive with an exponent $\beta \simeq 2.8$. The dashed line depicts the same case as the thick solid line, but with a $\Theta=10^{-4}$ thermal noise.}
	\label{LogofC2_a}
\end{figure}

In order to test Eq. \ref{see sub too}, we numerically solved  Eq. \ref{normalized ratchet equation} for 1,000 particles that were allowed to evolve from $z$=0 for a time $\tau$. The thick solid line in Fig. \ref{C2_lambda0p095_Gamma1p47} is from Eq. \ref{see sub too} and the thin lines are the results of numerical simulations. Eq. \ref{see sub too} gives a good fit to the data  in the interval before the saturation regime. In order to explore this region in more detail, in Fig. \ref{LogofC2_a} we show a log-log plot of $C_2$ as a function of $\tau$. The thick solid line is the result of the simulations with 1,000 particles and the thin solid line is the result of Eq. \ref{see sub too}.  For an extended range of values of $\tau$ the results indicate that $C_2$ grows with a power law of the form,
\begin{equation} \label{see sub too two}
C_2=A  \tau^\beta.
\end{equation}
In early times, particles are not yet trapped and the motion is ballistic with an exponent $\beta \approx 2$. In the intermediate times there are both trapped and running particles and over an extended time period, from $\tau\approx 10^2$ to $10^4$, the motion is superdiffusive with an exponent $\beta \simeq 2.8$. Beyond $\tau>10^4$, $C_2$ slowly levels off and crosses over to the saturation region, where particles are trapped. Our analytic result, Eq. \ref{see sub too}, is in excellent agreement with the data in the intermediate region, because we assume a continuous process that is only valid at longer time, before saturation.

We have obtained another estimate for $\beta$ from the expression for $C_2$, Eq. \ref{see sub too}, at  $\tau_1=1/(8 \alpha_T)$ and $\tau_2=1/(4 \alpha_T)$. For $\tau=\tau_1$ it is expected that 11.78\% of particles are trapped. For $\tau=\tau_2$ it is expected that 22.12\% of particles are trapped. Those $\tau$ values were chosen so that $C_2$ had not yet reached the saturation regime. From these considerations we can find $\beta$,
\begin{equation} \label{see sub too at tau sub one}
C_2(\tau_1)=A \enskip \tau_1^\beta=0.004069 \enskip \dfrac{v_{\omega}^{2} \enskip v_{r}^{2}}{\alpha_{T}^{2}}
\end{equation}
\begin{equation} \label{see sub too at tau sub two}
C_2(\tau_2)=A \enskip \tau_2^\beta=0.000575 \enskip \dfrac{v_{\omega}^{2} \enskip v_{r}^{2}}{\alpha_{T}^{2}}
\end{equation}
\begin{equation} \label{beta value}
\beta=\log_2 \bigg[ \dfrac{C_2(\tau_2)}{C_2(\tau_1)} \bigg]=2.82
\end{equation}
This result agrees well with the slope of the log-log plot in Fig. \ref{LogofC2_a}. Common logarithm is used. We note that the value of $\beta$ does not depend on $\alpha_T$ or $s$, provided there is just one velocity value for those particles that have not still been trapped. From Eq. \ref{see sub too} we see that there are two exceptions: $\alpha_T=0$ and $\alpha_T \rightarrow \infty$. In both cases $C_2=0$. When $\alpha_T=0$ there is no disorder at all. Then all particles move at the same velocity. When $\alpha_T \rightarrow \infty$ all particles get trapped in the first well.\\
The dashed line depicts the second cumulant when thermal noise is present. When particles get trapped in a well, they remain there, and therefore the no-thermal-noise curve tends towards a zero-slope asymptote, for the second cumulant grows no more when all particles have been trapped. However, with a low thermal noise of $\Theta=10^{-4}$, particles are trapped and also released continuously, because both the thermal noise and the quenched disorder are unbounded. Then, the second cumulant superdiffusive time extends for a longer time with the same $\beta$ exponent. The $\beta$ exponent for superdiffusion agrees with our prediction in this low noise situation because the only condition that we used to get it is that the trapping probability density is an exponential and, as stated before, this is met for low noise.  \\ 

\section{CONCLUSIONS}

In this paper, we have studied the role of correlated weak quenched disorder
on superdiffusive transport in a ratchet potential.  We have shown that the influence
of correlated quenched disorder allows  the presence of both running and trapped particles. Despite the fact that these processes
have only transient nature, they may last much longer than
characteristic time scales of the system and therefore be detectable experimentally.

In order to elucidate the relationship between anomalous superdiffusion and particle trapping  we focused on the particle trapping probability density function. Assuming that the trapping of particles is characterized by a constant probability of trapping per unit length  we analytically and numerically showed that the probability density function is a  decreasing exponential function,  with exponent $\alpha_P$ that characterizes the space dependence of the density function. We also show that the exponential function expression is valid for systems with low thermal noise.
Based on these results, we derive an analytical expression for the average number of wells where a given number of particles get trapped.

Finally, in an attempt to characterize superdiffusive motion, we obtain an analytic expression for  the second-moment distribution function $C_2$ as a function of time, when the particle velocity of particles that are not trapped is assumed to be constant.  Assuming that $C_2$ has a power law behavior with exponent $\beta$, we find a good fit to both our analytic expression for $C_2$ and the simulation results with $\beta=2.82$, indicating transient anomalous superdiffusive motion in the time period where trapping takes place.
\bigskip

\noindent {\bf ACKNOWLEDGEMENTS}

\bigskip
This work was partially supported by Universidad Nacional de Mar del Plata. CMA and FF acknowledge support from an American Physical Society International Travel grant that facilitated the continuation of the collaboration between CMA and FF.


\bibliographystyle{unsrt}
\bibliography{DGZbibratchet2}

\end{document}